\documentclass[aps,preprintnumbers,groupedaddress,longbibliography]{revtex4-2}

\usepackage{amssymb}
\usepackage{amsmath}
\usepackage{graphicx}
\usepackage{dcolumn}
\usepackage{newtxtext}
\usepackage{newtxmath}
\usepackage{natbib}
\usepackage{hyperref}
\usepackage{here}
\usepackage{bm}
\hypersetup{
    colorlinks = true,
    urlcolor   = blue,
    citecolor  = black,
}

\newcommand{\RomanNumeralCaps}[1]
\linenumbers
\usepackage{amsmath}

\usepackage{ulem}

\newcommand{\clmean}{C_{L,\textrm{mean}}}
\newcommand{\cdmean}{C_{D,\textrm{mean}}}
\newcommand{\dlmean}{\Delta_{L, \textrm{mean}}}
\newcommand{\ddmean}{\Delta_{D, \textrm{mean}}}
\newcommand{\Rey}{\textrm{Re}}
\newcommand{\Reyv}{\Rey_V}

\begin{document}

\title{Reynolds-number regimes of corrugation-induced lift enhancement in two-dimensional dragonfly-like wings}

\author{Yusuke Fujita}
\email[]{yuskfjt@riko.shimane-u.ac.jp}
\affiliation{Interdisciplinary Faculty of Science and Engineering, Shimane University, 1060, Nishikawatsu, Matsue, Shimane, 690-8504, Japan}

\author{Makoto Iima}
\email[]{iima@hiroshima-u.ac.jp}
\affiliation{Graduate School of Integrated Sciences for Life, Hiroshima University, 1-7-1, Kagamiyama, Higashi-Hiroshima, Hiroshima, 739-8521, Japan}

\date{\today}

\begin{abstract}
Corrugated wing structures commonly observed in insect wings can enhance aerodynamic performance at low Reynolds numbers ($\Rey \simeq O(10^3)$). However, their effectiveness strongly depends on the Reynolds number, and the corresponding operating range remains unclear. Direct numerical simulations of an impulsively started corrugated wing are performed over a wide range of Reynolds numbers ($100 \leq \Rey \leq 4000$) to identify the conditions under which corrugation is beneficial. Three  Reynolds-number regimes are identified within the examined parameter sets. No characteristic lift-enhancement mechanism is observed for $100 \leq \Rey < 1000$. For $1000 \leq \Rey \leq 4000$, lift enhancement is associated with alternating vortices formed within the V-shaped region of the corrugated wing, which generate low-pressure regions near the wing surface. Above $\Rey=2000$, the collapse and confinement of a secondary vortex within the V-shaped region emerge as an additional dominant mechanism, further enhancing the lift. An analysis based on a local Reynolds number organizes the onset conditions of these mechanisms and provides a useful local measure for characterizing the onset of vortex detachment from the corrugation. Furthermore, long-time simulations confirm that the observed vortex dynamics persist well beyond the initial transient response following an impulsive start. These results elucidate Reynolds-number-dependent lift-enhancement mechanisms and define aerodynamic conditions under which corrugated wings provide an advantage.
\end{abstract}

\maketitle

\section{\label{sec:1}Introduction}
Flying animals and vehicles come in various sizes, ranging from microscopic insects such as thrips to airplanes. Accordingly, the Reynolds number ($\Rey$) ranges from $O(10^1)$--$O(10^7)$ \citep{shyy2016aerodynamics, eldredge2019leading, liu2024vortices}, where $\Rey=\frac{cU_{\textrm{ref}}}{\nu}$, and $\nu$, $c$, and $U_{\textrm{ref}}$ represent the kinematic viscosity, wing chord length, and reference velocity (such as average flight speed), respectively. The lift-generation mechanism of an airplane at $\Rey \geq O(10^6)$ can be explained by the steady-wing theory based on the Kutta--Joukowski theorem \citep{landau2013fluid}. However, this theory cannot be applied to insects that achieve flight by flapping their wings at $\Rey = O(10^3)$~\citep{ellington1984aerodynamics}. Consequently, unsteady wing motion is a key factor in understanding insect flight.

When insects perform a flapping motion, they generate a leading-edge vortex (LEV) on the upper surface of the wing. LEVs have also been observed in bats and hummingbirds \citep{chin2016flapping}. The LEV contributes to lift enhancement caused by the low pressure at the vortex center. Therefore, LEV dynamics play a critical role in flapping flight at low Reynolds numbers \citep{shyy2016aerodynamics, ellington1996leading, sane2003aerodynamics, liu2024vortices}.

In addition to wing motion and Reynolds number effects, insects and airplanes have significantly different wing structures. The cross-sectional geometry of the wings of many insects such as dragonflies, bees, and cicadas can be described as a combination of V-shaped structures and other geometric features commonly categorized as corrugated wings \citep{dudley2002biomechanics}. These corrugated wings provide mechanical assistance during flapping \citep{bomphrey2016flight, newman1986approach, rajabi2016comparative} and contribute to sensing processes associated with foraging, escape, and travel \citep{bomphrey2016flight, uhrhan2024flow}. In addition, it has been postulated that this structure may offer aerodynamic advantages. For example, corrugated wings have the potential to exhibit high performance in the low-Reynolds-number regime ($\Rey \simeq O(10^{3})$) and are therefore of interest for designing small flying robots \citep{obata2014aerodynamic}.

Most previous studies that assessed the aerodynamic performance of corrugated wings evaluated their functionality under specific gliding conditions, typically at low angles of attack (AoA). For example, Obata and Sinohara demonstrated the superior flight stability of a corrugated wing at an AoA of $5^\circ$ for $\Rey \simeq 7000$ \cite{obata2009flow}. Kesel evaluated the aerodynamic performance of corrugated, flat, and filled wing models based on real dragonfly wings using wind-tunnel experiments at $\Rey \ge 10\,000$, reporting higher lift coefficients for corrugated wings corresponding to steady low-pressure regions formed in the surface valleys \cite{kesel2000aerodynamic}. Bomphrey et al. reported an improved lift-to-drag ratio at a low AoA of $\Rey=730$ \cite{bomphrey2016flight}. More recently, Narita and Chiba performed numerical simulations using three-dimensional scans of real wings of the wandering glider \textit{Pantala Flavescens} at $\Rey=4077$, suggesting that corrugation contributes to improved aerodynamic performance under gliding conditions \cite{narita2024aerodynamic}. In addition, Bauerheim and Chapin indicated that the unsteady flow induced by corrugation could become relevant even under gliding conditions at $\Rey = 6000$ \cite{bauerheim2020route} by investigating low-AoA cases. However, several studies conducted under low-AoA gliding conditions reported that the aerodynamic performance of corrugated wings is comparable to, or in some cases, lower than that of flat wings over a wide range of Reynolds numbers\citep{vargas2008computational, meng2011aerodynamic, meng2013aerodynamic, luo2005effects}. Thus, these studies suggest that the effectiveness of corrugation depends strongly on the flow regime and operating conditions.

Most conventional studies on corrugated wings have focused on static or quasi-steady characteristics that implicitly assume gliding flight \citep{rees1975aerodynamic, meng2011aerodynamic, meng2013aerodynamic, zhang2015aerodynamic, ansari2019optimal, zhan2025numerical}. However, these characteristics do not directly address the unsteady flow phenomena that arise in flapping flight, which is the typical flight mode of insects. Although directly reproducing realistic flapping kinematics is challenging due to the multiple coupled factors such as wing--body interaction, elasticity, and wake interaction, it is common in fluid dynamics research to extract key contributing elements and analyze them separately in simplified settings. Within this context, simplified two-dimensional models have been widely employed to investigate the fundamental aerodynamic mechanisms of insect-inspired wings, including corrugated geometries; these models can capture essential vortex dynamics under controlled conditions \citep{rival2014characteristic, zhang2015aerodynamic, ansari2019optimal}.

Despite the several numerical studies on corrugated wings under flapping conditions \citep{wang2022aerodynamic}, accurately isolating the aerodynamic role of corrugation from that of wing kinematics remains challenging. Consequently, many studies on biological and bio-inspired flight adopt approaches that decompose the flapping motion into elemental components (such as translation and rotation) and investigate each component separately \citep{eldredge2019leading}. Although the flow is inherently nonlinear and the effects of individual components cannot be simply superimposed, this decomposition can provide useful insights into dominant flow mechanisms in controlled numerical settings.

In a previous study, the authors considered the translational motion of a two-dimensional wing starting from rest, characterized by an impulsive initial condition \cite{fujita2023dynamic}, to clarify the relationship between the corrugated wing structure and lift generation associated with a component of unsteady motion. Even though the imposed translational motion is inherently steady, the resulting flow field exhibits unsteady vortex dynamics that govern lift generation. For a flat wing, a secondary vortex with a sign opposite that of the LEV, known as the $\lambda$ vortex, is generated near the wing surface \citep{eldredge2019leading}. This vortex is interpreted to repel the LEV and promote its release \citep{fujita2023dynamic, rival2014characteristic}.

For a corrugated wing, the authors suggested that the nonsmooth geometry can alter the interaction between the $\lambda$ vortex and the LEV and thereby improve aerodynamic performance through two mechanisms \cite{fujita2023aerodynamic, fujita2023dynamic}. The first mechanism involves the collapse and confinement of the $\lambda$ vortex within the V-shaped regions, which suppresses its negative influence on LEV retention and enables successive LEVs to remain closer to the wing, forming a low-pressure region. When this confinement is imperfect, the $\lambda$ vortex erupts into the bulk flow and contributes to LEV release, degrading performance. The second mechanism involves alternating vortices of opposite signs trapped within the V-shaped regions, which generate a strong negative pressure and further enhance lift. These vortex dynamics arise under high-AoA translational motion and differ from those observed for flat wings under the same conditions.

However, these previous studies were limited in their scope. The analysis was restricted to the short-time dynamics immediately following an impulsive start, and only a single Reynolds number ($\Rey=4000$) was considered. Long-time computations are required to assess whether the observed vortex dynamics persist and represent robust lift-enhancement mechanisms. Moreover, the Reynolds-number range over which these mechanisms are effective remains unclear. As the Reynolds number decreases, the viscous effects become dominant and the lift enhancement mechanisms likely vanish.

The Reynolds-number dependence of vortex-induced lift generation has been extensively studied in the context of flapping wings. For example, LEV shedding in the horizontal flapping of flat wings occurs only above a critical Reynolds number \citep{miller2004vortices}, and several studies have reported qualitative changes in vortex dynamics and lift generation across Reynolds-number ranges relevant to insect flight \citep{wang2000two, wang2007effect, tang2008effects, ota2012lift, harbig2013reynolds}. These results suggest that identifying the effective Reynolds-number range is essential for understanding unsteady lift mechanisms.

Therefore, in this study, we investigate the dependence of the aerodynamic performance of a corrugated wing undergoing impulsive translational motion on the Reynolds number. Numerical simulations are performed for $100 \leq \Rey \leq 4000$. For $100 \leq \Rey < 1000$, the flow fields around flat and corrugated wings are similar, and the aerodynamic performance is nearly independent of wing shape. For $1000 \leq \Rey < 2000$, lift enhancement correlates with alternating vortices within the V-shaped regions rather than $\lambda$-vortex collapse. For $2000 \leq \Rey \leq 4000$, the collapse of the $\lambda$ vortex becomes the dominant mechanism. These Reynolds-number regimes are characterised using the local Reynolds number associated with the V-shaped region of the corrugated wing $\Reyv$. Long-time simulations are also performed, demonstrating that the identified vortex dynamics persist beyond the initial transient and are not restricted to the early stages following an impulsive start.

\section{\label{sec:2}Method}
\subsection{\label{sec:2.1}Models}

\begin{figure}
	\centerline{\includegraphics{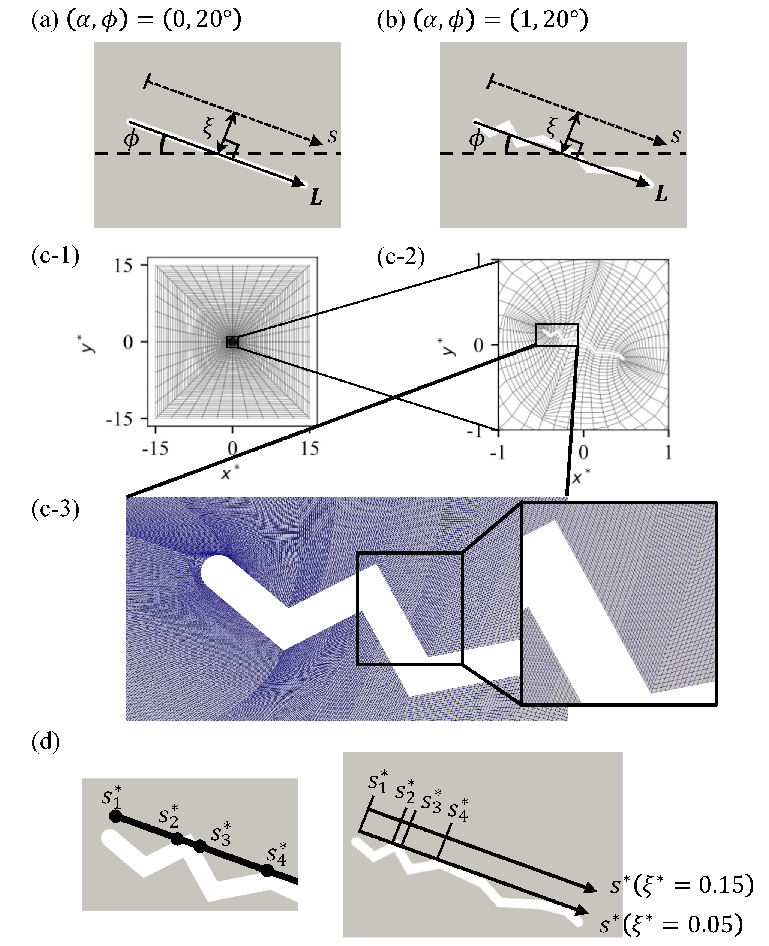}}
	\caption{\label{fig:1} Wing models. (a) A flat wing ($\alpha = 0$). (b) A corrugated wing ($\alpha = 1$). (c) Grid information ($\phi=20^\circ$). (c-1) Entire computational domain and spectral elements ($2960$ elements). (c-2) Computational domain and spectral elements around the corrugated wing model corresponding to the square in (c-1). (c-3) Distributions of collocation points around the tip of the corrugated wing model connected to adjacent points; the model corresponds to the rectangle in (c-2). The inset shows a magnified figure near a part of the corrugated wing model (corresponding to the rectangle in the main figure). (d) The positions, $s^*_1$, $s^*_2$, $s^*_3$, and $s^*_4$ for two cases of $\xi^*=0.05$ and $0.15$ with $(\alpha, \phi)=(1, 20^\circ$).}
\end{figure}

A corrugated wing model was constructed based on the wing of a southern hawker (\textit{Aeshna cyanea}) (Ref. \cite{kesel2000aerodynamic}, Profile 1). A two-dimensional wing model was generated based on a cross-section taken along a plane vertical to the line between the wing base and wing tip. This position is approximately 30\% of the wing base. Subsequently, a corrugated wing model was constructed by connecting these profile coordinates with line segments in sequence starting from the leading edge, and incorporating the thickness. We define a structural parameter $\alpha$, where $\alpha=0$ and $1$ correspond to the flat wings (Fig. \ref{fig:1} (a)) and the corrugated wing (Fig. \ref{fig:1} (b)), respectively. Furthermore, we define the vector $\mathbf{L}$ that connects the two ends of the connected line segments, which are the centers of the wing (from the leading to the trailing edge); the wing chord length $c=|\mathbf{L}|$ was $2$. The relative thickness of the wing $w/c$ was $0.04$, where $w$ represents the thickness of the line segments. The angle between the $\mathbf{L}$ and $x$ axes is defined as the AoA and denoted by $\phi$. Additional construction details are provided in Ref. \cite{fujita2023dynamic}.

\subsection{\label{sec:2.2} Numerical Simulation}
The two-dimensional incompressible Navier--Stokes equations nondimensionalized using the free-stream velocity $U_{\textrm{free}}$ and characteristic length $c$ are expressed as 

\begin{equation}
\frac{\partial \bm{u^*}}{\partial t^*} + (\bm{u^*} \cdot \nabla^*) \bm{u^*} = -\nabla^* p^* + \frac{1}{\Rey} \Delta^* \bm{u^*}, \quad \nabla^* \cdot \bm{u^*} = 0,
\label{eq:NS}
\end{equation}
where $\bm{u^*}=(u^*,v^*)$ and $p^*$ represent the velocity vector and pressure, respectively. The Reynolds number is defined by $\Rey=\frac{cU_{\textrm{free}}}{\nu}$, where $\nu$ represents the kinematic viscosity. Nondimensional variables are represented by the superscript $*$.

We used the spectral element method \citep{patera1984spectral} for numerical calculations \citep{fujita2023dynamic}, in which the computational domain was divided into a set of quadrilateral elements, and physical quantities in each element were independently represented by a polynomial function to ensure $C^0$ continuity across element boundaries \citep{karniadakis2005spectral}.

Consequently, the spectral element method balances the exponential (spectral) convergence of errors associated with global collocation methods such as pseudo-spectral methods with the geometric flexibility of traditional low-order finite element methods \citep{blackburn2019semtex}. The computational software \textit{Semtex} \citep{blackburn2019semtex} was used to solve Eq. (\ref{eq:NS}). 

The calculation methods are similar to those described in Ref. \cite{fujita2023dynamic}. The computational domain was $[-15,15] \times [-15,15]$ with the wing model centered at $(0,0)$. The outer edges of the computational domain, except for the right side ($x^*=15, -15 \le y^* \le 15$), were designated as inflow boundaries with $\bm{u^*}=\bm{U^*}\left(t^*\right)$, where $\bm{U^*}\left(t^*\right)=\left(0, 0\right)$ for $\left(t^*<0\right)$ and $\bm{U^*}\left(t^*\right)=\left(1, 0\right)$ for $\left(t^*>0\right)$, which enabled us to simulate an impulsive translation relative to a fixed wing. Outflow conditions on the right side were applied in the same manner as those described in Ref. \cite{fujita2023dynamic}. This investigation varies the values of $\nu$ to obtain Reynolds numbers in the range of $100 \leq \Rey \leq4000$. The results for $\Rey = 4000$ are identical to those obtained in our previous study \citep{fujita2023dynamic}. The time step $\Delta t^*$ was fixed at $2.0\times 10^{-5}$, and computations were performed within the range $0\leq t^* \leq 4$. Long-time computations were additionally performed for representative cases $(\Rey, \phi) = (4000, 20^\circ)$ and $(4000, 45^\circ)$ for both flat and corrugated wings to assess the robustness of the lift-generation mechanisms identified in Ref. \cite{fujita2023dynamic}.

Figures \ref{fig:1} (c-1) and (c-2) show the spectral elements surrounding the corrugated wing model at the AoA of $\phi=20^\circ$. The left side shows spectral elements of the total computational domain ($[-15, 15] \times [-15, 15]$) and the right side shows those around the corrugated wing model ($[-1, 1] \times [-1, 1]$). The total number of spectral elements was $N=2960$, with each spectral element discretized using an $N_p \times N_p = 19 \times 19$ mesh. Therefore, $N_p$ represents the number of colocation points along the edge of each element. Therefore, the number of colocation points is different from the number of spectral elements. Figure \ref{fig:1} (c-3) shows colocation points around the corrugated wing model. The validity of the computational grid and computational parameters was verified for $\Rey=4000$, $\phi=20^\circ$ and $35^\circ$ using the following comparisons: $(N_p, \Delta t^*)=(19, 5.0\times 10^{-5})$, $(19, 1.5\times 10^{-5})$, $(21, 2.0\times 10^{-5})$, $(15, 2.0\times 10^{-5})$, and $(7, 2.0\times 10^{-5})$, which were already examined in Ref. \cite{fujita2023dynamic}. Further, we compared the results for $\dlmean$ obtained using different computational domains $[-15, 15] \times [-15, 15]$ in Ref. \citep{fujita2023dynamic} and $[-5, 5] \times [-5, 5]$ in Ref. \citep{fujita2023aerodynamic}. The results demonstrated good agreement, suggesting that the computational domain used in this study was sufficiently large and the effect of the domain boundaries was minimal.

\subsection{\label{sec:2.3} Evaluation of Aerodynamic Performance}
The aerodynamic performance was evaluated using the lift coefficient $C_L(\alpha, \phi, \Rey; t^*)=L/[\left(1/2\right)\rho cU_{\textrm{free}}^{2}]$, where $L$ represents the lift, and the drag coefficient $C_D(\alpha, \phi, \Rey; t^*)=D/[\left(1/2\right)\rho cU_{\textrm{free}}^{2}]$, where $D$ represents the drag. The mean lift coefficients within the time interval $[t^*_1, t^*_2]$ denoted by $C_{L,\textrm{mean}}(\alpha, \phi, \Rey)$ are defined as 
\begin{equation}
	C_{L,\textrm{mean}}(\alpha, \phi, \Rey)=\frac{1}{t^*_2-t^*_1} \int_{t^*_1}^{t^*_2} C_L(\alpha, \phi, \Rey;t^*) dt^*,
	\label{eq:def mean}
\end{equation}
with similar definitions for $\cdmean$.
The relative difference in $\clmean$ between the corrugated ($\alpha =1$) and flat wings ($\alpha =0$), $\dlmean$, is defined as 
\begin{equation}
	\dlmean(\phi, \Rey) = \frac{\clmean(1, \phi, \Rey)-\clmean(0,\phi, \Rey)}{\clmean(0,\phi, \Rey)},
	\label{eq:dlmean}
\end{equation}
with similar definitions for $\ddmean$.
In this study, $\dlmean$ was used as an indicator of wing performance. If $\dlmean = 0$, the wing performances of the corrugated and flat wings are comparable. When $\dlmean > 0$, the wing performance of the corrugated wing is better than that of the flat wing.

The pressure distribution was quantified using the pressure coefficient defined as $C_p = \frac{p - p_\infty}{\tfrac{1}{2}\rho |\bm{U}|^2}$, where $p_\infty$ represents the inflow pressure measured at $y^* = -15$. Vortices directly generate low-pressure regions and contribute to lift enhancement at $\Rey=4000$ \citep{fujita2023dynamic}. This relationship is clearly demonstrated in Figs. 6--8, 10, and 14 of that study. In the present study, pressure-based analyses were restricted to long-time simulations. For the other cases, we focused on vorticity fields and investigated the Reynolds-number dependence of the relationship between vortex dynamics and lift-enhancement mechanisms. Vorticity is defined as $\omega_z^* = \partial v^*/\partial x^* -  \partial u^*/\partial y^*$. We analyzed the spatiotemporal distribution of vorticity in a region near the upper surface of the wing model characterized by the distribution on a line segment defined by a shifted vector of $\mathbf{L}$ by $\mathbf{\xi}$, where $\mathbf{\xi} \perp \mathbf{L}$ (as shown in Figs. \ref{fig:1} (a) and (b)). The position along this line segment is denoted by $s$. Both the shift magnitude $|\mathbf{\xi}|$ and parameter $s$ are normalized by the chord length of the wing, leading to dimensionless parameters $\xi^*=\frac{|\mathbf{\xi}|}{c}$ and $s^*=\frac{s}{c}$. When $\xi^* =0$, $s^*=0$ and $1$ correspond to the leading and trailing edges, respectively. 

In this study, we examined flow characteristics at $\xi^*=0.05$ and $0.15$ based on specific flow features discussed in the following section. To examine flow characteristics near the leading edge (Fig. \ref{fig:1} (d)), we define $(s^*_1, s^*_2, s^*_3, s^*_4)=(0.000, 0.144, 0.188, 0.353)$ and use these values in the discussion (Section \ref{sec:3.4}--Section \ref{sec:3.6}). Figure \ref{fig:1} (d) shows these positions $s^*_1$, $s^*_2$, $s^*_3$, and $s^*_4$ for $(\alpha, \phi)=(1, 20^\circ)$ along $\xi^*=0.05$ and $0.15$. For $\xi^*=0.05$, $s^*_2$, $s^*_3$, and $s^*_4$ represent the intersection points of the $s^*$-axis with the wing model; the same values were used for $\xi^* = 0.15$.

We focus on the period $[t^*_1, t^*_2]=[0.50, 3.25]$ (see Fig. \ref{fig:2}). This period corresponds to the development of the LEV and $\lambda$ vortex generated on the upper wing surface, including vortex collapse and interactions before vortices detach near the wing \citep{fujita2023dynamic}. For $\Rey=4000$, fixing $t^*_2 = 3.25$ and varying $t^*_1$ within the range $0.02\leq t^*_1\leq1.10$ do not qualitatively affect the results, demonstrating that the present conclusions are robust to moderate variations in the onset timing of the period \citep{fujita2023dynamic}.

We extract a fundamental component of wing kinematics, i.e., translational motion initiated from rest to simplify the overall flapping motion. The time interval considered in the simplified wing motion model approximately corresponds to a specific phase of the actual flapping cycle of the dragonfly. Southern hawker wing (\textit{Aeshna cyanea}, Profile 1 in Ref. \cite{kesel2000aerodynamic}) was examined at a spanwise position 1.50 cm from the wing root, where the chord length was 0.74 cm (based on scanned data). Wing kinematics were characterized by a wingbeat frequency of 36 Hz and a stroke amplitude of $50^\circ$, as reported by Ref. \cite{ruppell1989kinematic}. Based on these values, the flapping velocity at the specified spanwise location was estimated to reach a maximum value of 148 cm/s. Consequently, the analysis period $[t^*_1, t^*_2]=[0.50, 3.25]$ corresponded to a dimensional time interval of approximately 0.014 s. This interval represented approximately 50\% of a complete wingbeat cycle. Assuming that the upstroke and downstroke occupied equal durations, this time window coincided with the downstroke phase during which substantial lift was generated.

\newpage
\section{\label{sec:3} Results and discussion}

\subsection{\label{sec:3.1}Reynolds number dependency of $C_L$}

\begin{figure}
	\centerline{\includegraphics{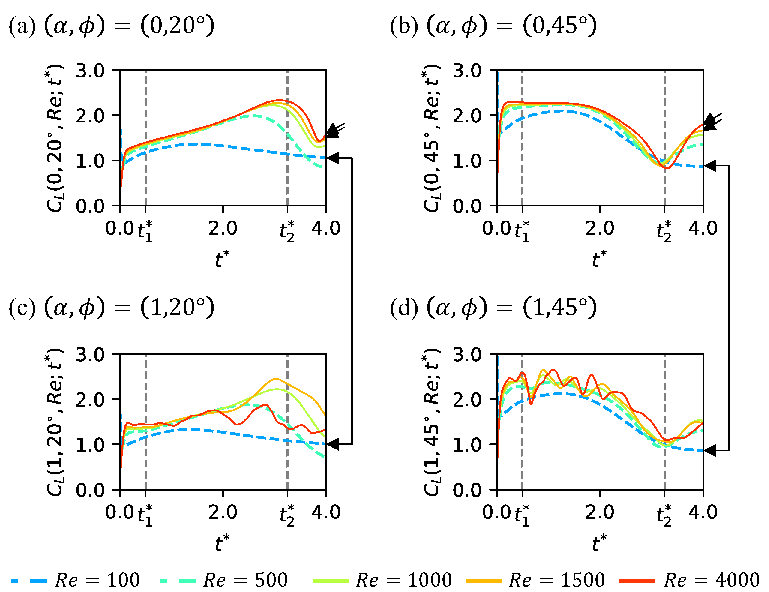}}
	\caption{\label{fig:2} Time series of $C_L (\alpha, \phi, \Rey; t^*)$ for the Reynolds numbers, respectively. (a) $(\alpha,  \phi)=(0, 20^\circ)$, (b) $(\alpha,  \phi)=(0, 45^\circ)$, (c) $(\alpha,  \phi)=(1, 20^\circ)$, and (d) $(\alpha,  \phi)=(1, 45^\circ)$.}
\end{figure}

Figure \ref{fig:2} shows the time series of the lift coefficients of flat and corrugated wings for $\phi=20^\circ$ and $45^\circ$.

In the case of $\Rey \leq 500$ (dashed lines), the shapes of $C_{L}(0, \phi, \Rey; t^*)$ and $C_{L}(1, \phi, \Rey; t^*)$ are almost identical for both values of $\phi$.

However, when $\Rey \geq 1000$, the shapes of $C_{L}(\alpha, \phi, \Rey; t^*)$ for $\alpha = 0$ and $1$ differ significantly. For a flat wing, the shapes of $C_{L}(0, \phi, \Rey; t^*)$ for $\Rey = 1500$ and $4000$ are similar for both values of $\phi$ (Figs. \ref{fig:2} (a) and (b)), which is consistent with the results of previous experimental studies \citep{mueller1999aerodynamic, okamoto2016effectiveness} in which the wing performance (lift coefficient) of a flat wing did not change when $\Rey=O(10^3)$. For the corrugated wing, there was a significant difference between $C_{L}(1, \phi, 1500; t^*)$ and $C_{L}(1, \phi, 4000; t^*)$ (Figs. \ref{fig:2} (c) and (d); orange and red lines).

\newpage
\begin{figure}
	\centering{\includegraphics{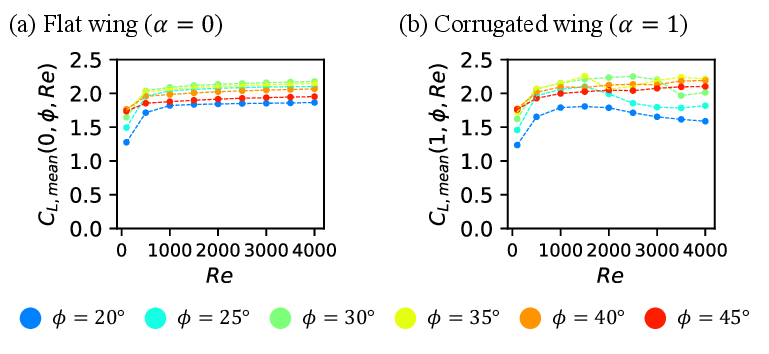}}
	\caption{\label{fig:3}Reynolds number dependency of $\clmean(\alpha, \phi, \Rey)$ for $\phi$, respectively. (a) The flat wing ($\alpha=0$) and (b) corrugated wing ($\alpha=1$).}
\end{figure}

Figure \ref{fig:3} shows the mean lift coefficient $\clmean(\alpha, \phi, \Rey)$.

For $\Rey \leq 500$, $\clmean(\alpha, \phi, \Rey)$ was similar for both wings, which agrees with the fact that $C_{L}(\alpha, \phi, \Rey; t^*)$ did not differ between the wings (Fig. \ref{fig:2}). This suggests that the corrugated wing structure does not affect wing performance in this Reynolds number regime.

For $\Rey \geq 1000$, $\clmean(\alpha, \phi, \Rey)$ for the flat wing was almost constant for all Reynolds numbers (Fig. \ref{fig:3} (a)). This result is consistent with that of the flapping wing reported in Ref. \cite{wang2007effect}. However, for a corrugated wing, $\clmean(\alpha, \phi, \Rey)$ is not constant (Fig. \ref{fig:3} (b)). When the Reynolds number increases, the value of $\clmean(\alpha, \phi, \Rey)$ decreases slightly ($\phi=20^\circ$; blue line) or varies noticeably ($\phi=25^\circ$ and $30^\circ$; Fig. \ref{fig:3} (b)). In terms of $\clmean$, the effect of the corrugated wing structure is apparent above $\Rey =1000$.

\newpage
\subsection{\label{sec:3.2}Reynolds number dependency of $\dlmean$ and $\ddmean$}

\begin{figure}
	\centering{\includegraphics{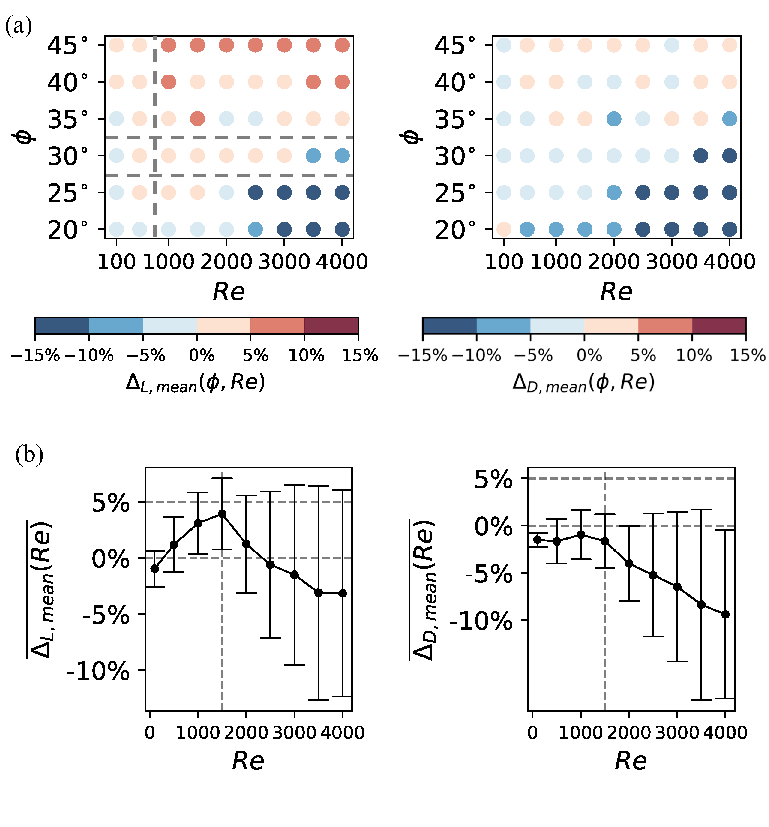}}
	\caption{\label{fig:4} (a) $\dlmean(\phi, \Rey)$ and $\ddmean(\phi, \Rey)$ for the Reynolds numbers and $\phi$s. (b) Reynolds number dependency of the average of $\dlmean(\phi, \Rey)$ and $\ddmean(\phi, \Rey)$ for $\phi$. Error bars indicate standard deviation.}
\end{figure} 

A more detailed comparison of the aerodynamic performance of the flat and corrugated wings is shown in Fig. \ref{fig:4} (a), which illustrates the relative aerodynamic performance of the corrugated wing $\dlmean(\phi, \Rey)$. When $\phi > 30^\circ$, $\dlmean(\phi, \Rey)$ is positive in almost all cases, indicating an improved aerodynamic performance for the corrugated wing. However, when $\phi < 30^\circ$, $\dlmean(\phi, \Rey)$ is negative in almost all cases, showing no improvement in aerodynamic performance. Therefore, the transition in $\dlmean(\phi, \Rey)$ was approximately $\phi=30^\circ$; the same applies to drag. The transition of $\ddmean(\phi, \Rey)$ was made around $\phi=30^\circ$. However, when $\phi > 30^\circ$, $\ddmean(\phi, \Rey)$ does not increase significantly. Thus, the corrugated wing can enhance lift without incurring a drag penalty. Consequently, from this point on, we focus solely on the lift and discuss the mechanism of vortices that increase the lift.

For $\Rey < 1000$, the absolute value of $\dlmean(\phi, \Rey)$ is small ($|\dlmean(\phi, \Rey)|<0.04$) independent of $\phi$, which indicates no significant difference in performance. However, as the Reynolds number increases, the absolute value of $\dlmean(\phi, \Rey)$ tends to increase, confirming the difference in aerodynamic performance. For example, at $\phi = 45^\circ$, $\dlmean(45^\circ, \Rey) < 0.05$ for $\Rey \leq 500$; however, $\dlmean(45^\circ, \Rey) > 0.05$ for $\Rey \geq 1000$. In contrast, at $\phi = 20^\circ$, $\dlmean(20^\circ, \Rey) > -0.05$ for $\Rey \leq 2000$, $\dlmean(20^\circ, \Rey) < -0.05$ for $\Rey \geq 2500$. The results indicate that the Reynolds number at which performance differences between the flat and corrugated wings become significant depends on $\phi$.

We aim to simplify the understanding of trends in the Reynolds numbers. For each Reynolds number, the mean value of $\dlmean(\phi, \Rey)$ for $20^\circ \leq \phi \leq 45^\circ$ was defined as $\overline{\dlmean(\Rey)}$. Figure \ref{fig:4} (b) shows $\overline{\dlmean(\Rey)}$. For $100 \leq \Rey \leq 1500$, $\overline{\dlmean(\Rey)}$ increases with the Reynolds number. However, for $1500 \leq \Rey \leq 4000$, $\overline{\dlmean(\Rey)}$ decreases with an increase in the Reynolds number. Among the Reynolds numbers examined, the angle-averaged relative lift difference attains its maximum at $Re=1500$, which is consistent with the findings of Ref. \cite{fujita2023aerodynamic}. The error bars in Fig. \ref{fig:4} (b) represent the standard deviation (SD). The smallest SD was observed at $\Rey = 100$. For $\Rey \leq 3500$, the SD increased monotonically with the Reynolds number, and the SD at $\Rey = 4000$ was slightly smaller but similar to that at $\Rey = 3500$. A similar trend was observed for drag, represented by $\overline{\ddmean(\Rey)}$.

\newpage
\subsection{\label{sec:3.3}Vortex pattern}
\subsubsection{\label{sec:3.3.1} In the case of $\phi=45^\circ$}

\begin{figure}
	\centering{\includegraphics{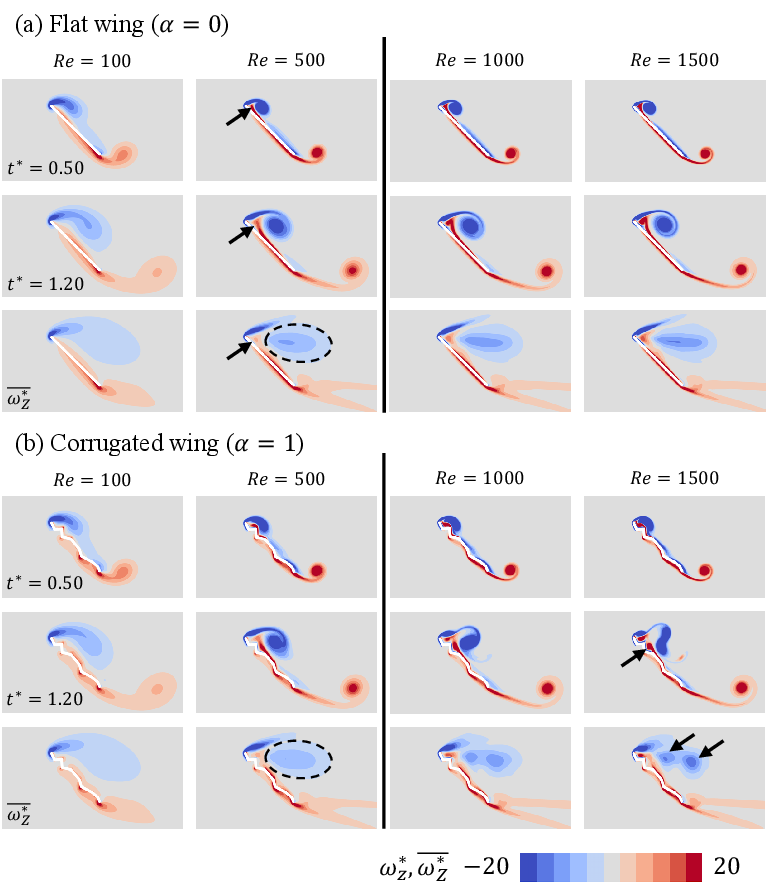}}
	\caption{\label{fig:5} Vorticity fields around wings for $\phi=45^\circ$ at $\Rey=100$, $500$, $1000$, and $1500$. The first and second rows display snapshots at $t^* = 0.50$ and $t^* = 1.20$, respectively. The third row shows the mean vorticity field, $\overline{\omega_z^*}$ in $[t^*_1, t^*_2]$. (a) The flat wing ($\alpha=0$) and (b) corrugated wing ($\alpha=1$).}
\end{figure}

Vorticity fields around the wings for $\phi=45^\circ$ at $\Rey=100$, $500$, $1000$, and $1500$ are shown in Fig. \ref{fig:5} to compare flow patterns associated with the improved aerodynamic performance of the corrugated wing. Figures \ref{fig:5} (a) and (b) show vorticity fields around flat and corrugated wings, respectively. The first and second rows show snapshots at $t^* = 0.50$ and $t^* = 1.20$, respectively. The third row presents the mean vorticity field $\overline{\omega_z^*}$ over $[t^*_1, t^*_2]$.

For the flat wing (Fig. \ref{fig:5} (a)), the vortex at $t^* = 0.50$ grew in a self-similar manner independent of the Reynolds number. At $\Rey=100$, a large region of negative vorticity corresponding to the LEV forms above the wing. As the Reynolds number increased, the LEV became more pronounced and exhibited a circular structure. Adjacent to the LEV, a vortex with positive vorticity identified as the $\lambda$ vortex was observed (see the arrows for $\Rey=500$). This $\lambda$ vortex grows and erupts between two LEVs when one LEV is released \citep{rival2014characteristic, eldredge2019leading, fujita2023dynamic}.

For mean vorticity fields at $\Rey=500$, $1000$, and $1500$, an elongated region of negative vorticity was observed along the upper surface of the wing (dashed ellipse for $\Rey=500$). This elongation resulted from the advection of the LEV downstream. The magnitude of this negative vorticity increased monotonically with the Reynolds number. Furthermore, a positive vorticity distribution was observed between the leading-edge negative vorticity and elongated negative vorticity region on the upper surface; this was attributed to the eruption of the $\lambda$ vortex (see arrow for $\Rey=500$).

For the corrugated wing (Fig. \ref{fig:5} (b)), the vortex at $t^* = 0.50$ grew in a self-similar manner at $\Rey=100$ and $500$, similar to that in the flat wing case. However, at $\Rey=1000$ and $1500$, vortex growth did not exhibit self-similarity. The positive vortex splits into multiple smaller vortices, which become stuck in the V-shaped regions (arrow for $\Rey=1500$), confirming the collapse of the $\lambda$ vortex because of the corrugated structure. In addition, two LEVs were aligned side-by-side above the wing, consistent with observations at $\Rey=4000$ \citep{fujita2023dynamic}.

For the mean vorticity field at $\Rey=500$, an elongated negative vorticity distribution appears along the upper surface (dashed ellipse) similar to that in the flat-wing case (Fig. \ref{fig:5} (a)). However, at $\Rey=1000$ and $1500$, two distinct negative-vorticity regions were formed above the wing (arrows for $\Rey=1500$) because the fully separated LEV moved above the corrugated wing in an isolated state.

Comparing the results for both wings (Figs. \ref{fig:5} (a) and (b)) revealed that vortex dynamics around the corrugated wing differs from those around the flat wing above $\Rey \geq 1000$ with an increase in the Reynolds number. The transition of vortex dynamics in the corrugated wing improved the aerodynamic performance of the corrugated wing (Fig. \ref{fig:4}). The collapse of the $\lambda$ vortex and interaction between the two LEVs caused the LEV to move closer to the corrugated wings. The vortex center corresponds to the negative-pressure region, and therefore, the negative pressure distribution on the corrugated wing surface becomes wider than that on the flat wing (data not shown; refer to Ref. \cite{fujita2023dynamic}). Consequently, the aerodynamic performance of the corrugated wing was better than that of the flat wing. Thus, vortex dynamics are strongly associated with the lift enhancement mechanism of corrugated wings.

\newpage
\subsubsection{\label{sec:3.3.2} In the case of $\phi=20^\circ$}

\begin{figure}
	\centering{\includegraphics{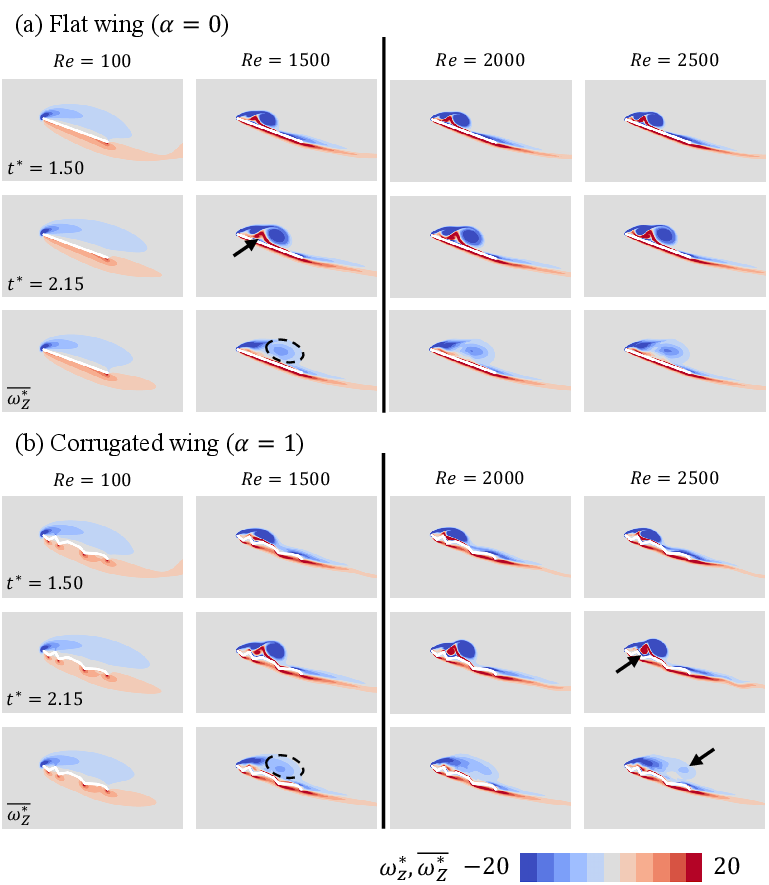}}
	\caption{\label{fig:6} Vorticity fields around wings for $\phi=20^\circ$ at $\Rey=100$, $1500$, $2000$, and $2500$. The first and second rows display snapshots at $t^* = 1.50$ and $t^* = 2.15$, respectively. The third row shows the mean vorticity field $\overline{\omega_z^*}$. (a) The flat wing ($\alpha=0$) and (b) corrugated wing ($\alpha=1$).}
\end{figure}

Vorticity fields around the wings at $\phi=20^\circ$ and $\Rey=100$, $1500$, $2000$, and $2500$ are shown in Fig. \ref{fig:6} to compare flow patterns in cases where the aerodynamic performance of the corrugated wing did not improve. Figures \ref{fig:6} (a) and (b) show vorticity fields around the flat and corrugated wings, respectively. The first and second rows show snapshots at $t^* = 1.50$ and $t^* = 2.15$, respectively, whereas the third row shows the mean vorticity field $\overline{\omega_z^*}$ over $[t^*_1, t^*_2]$.

For the flat wing (Fig. \ref{fig:6} (a)), the vortex at $t^* = 1.50$ exhibits self-similar growth across all Reynolds numbers, similar to the case of $\phi=45^\circ$ (Fig. \ref{fig:5} (a)). At $\Rey=100$, a large region of negative vorticity corresponding to LEV forms above the wing. As the Reynolds number increases, the LEV becomes more pronounced and maintains its circular structure. A positive vorticity region identified as the $\lambda$ vortex appeared adjacent to the LEV (arrow at $\Rey=1500$).

For the mean vorticity field at $\phi=20^\circ$, the flow patterns are similar to those at $\phi=45^\circ$ (Fig. \ref{fig:5} (a)). However, unlike the $\phi=45^\circ$ case, the LEV moved near the wing surface, preventing a distinct $\lambda$ vortex distribution. This led to a more concentrated negative vorticity region near the wing.

For the corrugated wing (Fig. \ref{fig:6} (b)), the vortex at $t^* = 1.50$ grew in a self-similar manner to that at $\Rey=100$, similar to the flat wing. At $\Rey = 1500$, although the vortices at $t^* = 1.50$ are not similar to those of the flat wing, the shape of the $\lambda$ vortex at $t^* = 2.15$ is similar. Therefore, we consider the vortex to grow in a self-similar manner (refer to the mean vorticity field and Fig. \ref{fig:8}). However, at $\Rey=2000$ and $2500$, vortex growth did not exhibit self-similarity. At $t^* = 2.15$, a positive vortex with a more rounded shape was observed adjacent to the two LEVs (arrow at $\Rey=2500$), indicating a $\lambda$-vortex eruption caused by the corrugated structure. This phenomenon is observed for $\Rey\geq2000$ and is consistent with observations at $\Rey=4000$ \citep{fujita2023dynamic}.

For the mean vorticity field at $\Rey=100$ and $1500$, the flow patterns are similar to those obtained in the flat-wing case (Fig. \ref{fig:6} (a)). However, at $\Rey=2000$, this distribution is not present. At $\Rey=2500$, a single distinct negative vorticity region is observed above the wing (arrow) in contrast to the two separate regions observed at $\phi=45^\circ$ (Fig. \ref{fig:5} (b)). This confirms that the $\lambda$-vortex eruption caused the LEV to move further away from the wing, weakening the vorticity distribution \citep{fujita2023dynamic}.

Comparing both wings (Figs. \ref{fig:6} (a) and (b)) reveals that vortex dynamics around the corrugated wing deviate from those of the flat wing in $\Rey > 1500$. Although $|\dlmean(20^\circ, 2000)|=|-0.031|<0.05$ (Fig. \ref{fig:4} (a)), the flow begins to exhibit changes at $\Rey=2000$. The $\lambda$ vortex erupted with an increase in the Reynolds number, causing the LEV to move away from the corrugated wing. This reduces the lift enhancement mechanism observed at $\phi=45^\circ$ and results in no significant improvement in the aerodynamic performance \citep{fujita2023dynamic}.

\newpage
\subsubsection{\label{sec:3.3.3} Spatiotemporal distribution of $\omega_z^*$ on $\xi^* = 0.15$}

\begin{figure}
	\centering{\includegraphics{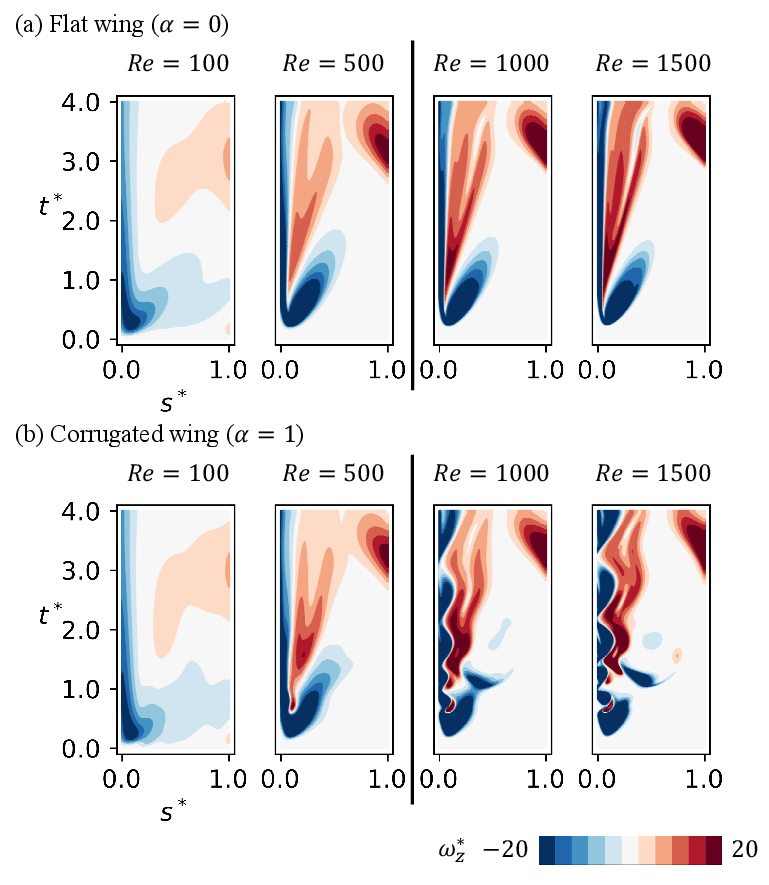}}
	\caption{\label{fig:7} Spatiotemporal distribution of $\omega_z^*$ on $\xi^* = 0.15$ for $\phi = 45^\circ$. The conditions are the same as that in Fig. \ref{fig:5}.}
\end{figure}

The spatiotemporal flow pattern of the flat wing was compared with that of the corrugated wing for different Reynolds numbers ($\phi=45^\circ$). Figure \ref{fig:7} shows the spatiotemporal distribution of the vorticity along the $s^*$ axis at $\xi^* = 0.15$. Figures \ref{fig:7} (a) and (b) show the results for flat and corrugated wings, respectively. At $\Rey=100$ and $500$, the vortex pattern of the flat wing was almost the same as that of the corrugated wing. At $\Rey=1000$ and $1500$, the vortex patterns of these wings differ substantially in the region near the leading edge (small values of $s^*$); an oscillation is observed (e.g., Fig. \ref{fig:7} (b); $\Rey=1500$). We focused on the distribution of $\xi^* = 0.15$ at a distance from the V-shaped region; therefore, the alternation of vortices was represented by an oscillation.

\newpage
\begin{figure}
	\centering{\includegraphics{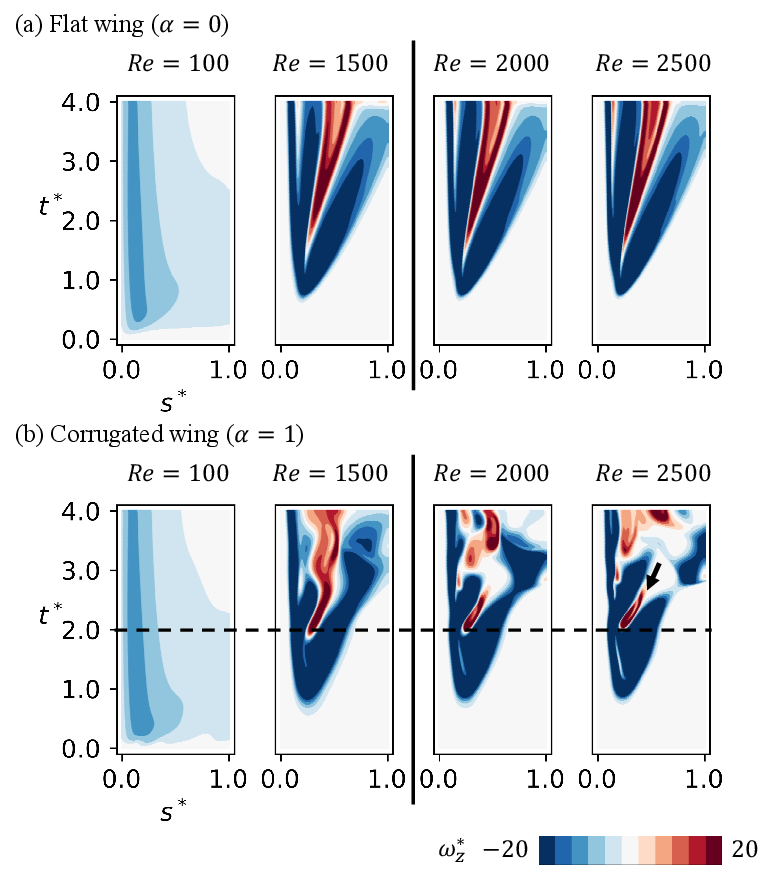}}
	\caption{\label{fig:8} Spatiotemporal distribution of $\omega_z^*$ on $\xi^* = 0.15$ for $\phi = 20^\circ$. The conditions are the same as that in Fig. \ref{fig:6}.}
\end{figure}

Figures \ref{fig:8} (a) and (b) show the spatiotemporal distribution of vorticity along the $s^*$ axis at $\xi^* = 0.15$ ($\phi=20^\circ$) for flat and corrugated wings, respectively. At $\Rey=100$, the vortex patterns of both wings were similar to those at $\phi=45^\circ$. At $\Rey=1500$, the vortex pattern of the flat wing was qualitatively similar to that of the corrugated wing; however, a slight oscillation was observed in the positive vorticity region. At $\Rey = 2000$ and $2500$, the vortex patterns on the wings differed substantially. In addition, the $\lambda$-vortex eruption is represented as a vorticity region with a positive sign (red) around $(s^*, t^*) = (0.25, 2)$ at $(\alpha, \phi, \Rey) = (1, 20^\circ, 2500)$, which indicates advection towards the trailing edge (in the direction of increasing $s^*$) (Fig. \ref{fig:8} (b); $\Rey=2500$, arrow). However, such features are not observed when the wing performance improved (Fig. \ref{fig:7} (b); $\Rey=1500$). The detailed mechanism is discussed in Ref. \cite{fujita2023dynamic}.

\subsection{\label{sec:3.4} Flow characterization near the V-shaped region}

\begin{figure}
	\centering{\includegraphics{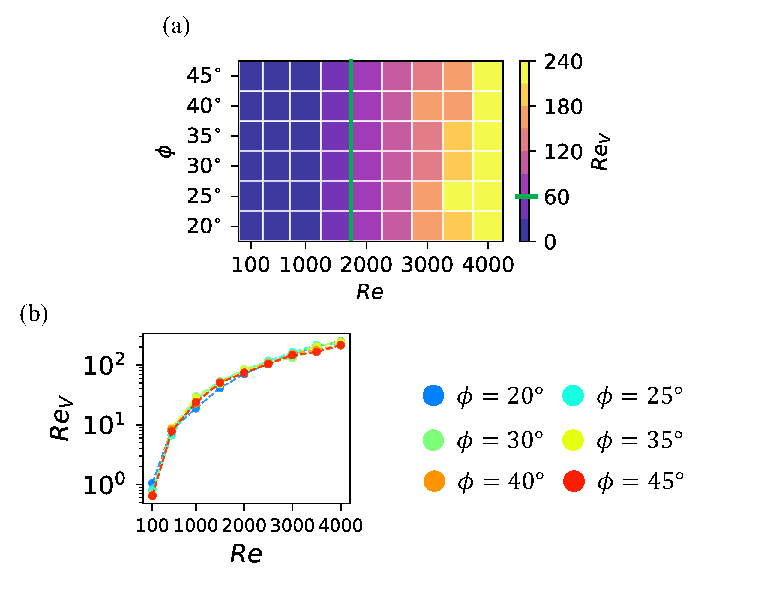}}
	\caption{\label{fig:9} (a) $\Reyv$ for $20^\circ\leq\phi\leq45^\circ$ and $100\leq\Rey\leq4000$. (b) Reynolds number dependency of $\Reyv$ for $\phi$, respectively. We used a log scale on the vertical axis.}
\end{figure}

As indicated in Section \ref{sec:3.3}, the characteristic behaviour of vortex motion on the flat wing and that on the corrugated wing differ with increasing Reynolds number. We hypothesize that this change is attributable to flow characteristics near the concave-convex structure of the corrugated wing. To quantify this, we defined the local Reynolds number characterized by the V-shaped structure of the corrugated wing $\Reyv$ as

\begin{equation}
	\Reyv:=\frac{bU_V}{\nu},
\label{eq:3.1}
\end{equation}

where $b$ represents the characteristic depth of the V-shaped region and $U_{V}$ represents the characteristic flow speed in the V-shaped region. We set $b=\frac{2}{9}$ as the aspect ratio of the corrugated wing model (Fig. \ref{fig:1} (b)), which can be estimated approximately as $9:1$. The value of $U_{V}$ was estimated from the average velocity along the line at $\xi^*=0.05$ as

\begin{equation}
	\begin{split}
	U_s(t^*)&=\frac{1}{({s^*_2}-{s^*_1})+({s^*_4}-{s^*_3})} \left\{\int_{s^*_1}^{s^*_2} |\boldsymbol{U}(s^*, t^*)| ds^*+\int_{s^*_3}^{s^*_4} |\boldsymbol{U}(s^*, t^*)| ds^*\right\},\\
	U_V&=\frac{1}{t^*_2-t^*_1} \int_{t^*_1}^{t^*_2} U_s(t^*) dt^*.
	\end{split}
\label{eq:3.2}
\end{equation}

Here, the interval $s^*_2 < s^* < s^*_3$ is excluded because it overlaps with the convex part (Fig. \ref{fig:1} (d)).

Figure \ref{fig:9} (a) shows $\Reyv$ for $(\phi, \Rey)$. $\Reyv$ increases with an increase in the Reynolds number; however, it becomes less dependent on $\phi$. In particular, $\Reyv\geq60$ for $2000\leq\Rey\leq4000$ (Fig. \ref{fig:9} (a); green line). In the range of $2000\leq\Rey\leq4000$, flow fields around the corrugated wing were affected by the corrugated structure, resulting in a deviation in the performance from that of the flat wing; however, details depend on the value of $\phi$ (Figs. \ref{fig:4} (b), \ref{fig:7}, and \ref{fig:8}). According to the observations, the critical $\Reyv$ in this system is $60$, which approximately corresponds to the vortex detachment of other flapping-wing models (\cite{miller2004vortices} and \cite{ota2012lift}). This agrees with our observation that vortex detachment ($\lambda$-vortex eruption or collapse) from the tip of the corrugated structure occurs at $60\leq\Reyv$ ($2000\leq\Rey$).

Figure \ref{fig:9} (b) shows the Reynolds-number dependence of $\Reyv$ for $\phi$, respectively. $\Reyv$ is a function of $\Rey$ irrespective of $\phi$. We observe that this is $\Reyv<10$ for $100\leq\Rey\leq500$, indicating that the flow around the corrugated wing for $100\leq\Rey\leq500$ is not significantly affected by the concave-convex structure, which agrees with our observations.

A decrease in the Reynolds number increases the effect of viscosity, and vortex shedding from the convex parts of the corrugated wing is no longer observed. Consequently, vortex dynamics above the corrugated wing became similar to those over the flat wing. Consequently, no significant changes in wing performance were observed.

\subsection{\label{sec:3.5} Vortex oscillations associated with lift enhancement}

\begin{figure}
	\centering{\includegraphics{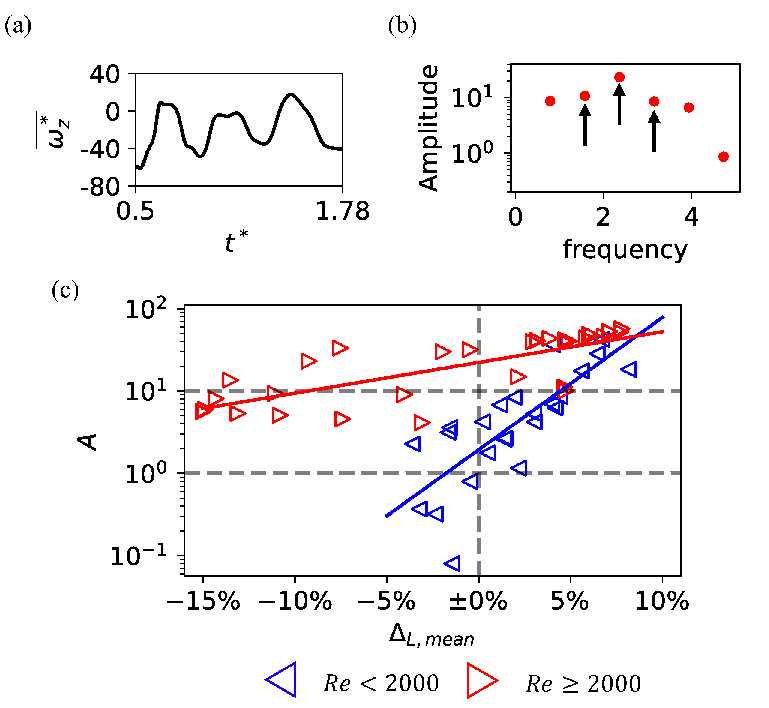}}
	\caption{\label{fig:10} (a) Time series of $\overline{\omega^*_z}$ for $(\alpha, \phi, \Rey)=(1, 45^\circ, 1500)$. (b) Results of Fourier analysis for (a). (c) Scatter plot of $\dlmean$ and amplitude of specific frequencies, $A$. Blue and red triangles represent the cases of $\Rey <2000$ and $\Rey \geq 2000$, respectively. The straight line results from a regression curve for each Reynolds number regime.}
\end{figure}
	
As the Reynolds number increased, the flow around the corrugated wing was influenced by the corrugated structure, resulting in the oscillation of the vortex in the V-shaped region, as indicated in Fig. \ref{fig:7} (b) at $\Rey=1500$. Consequently, the oscillation enhances the aerodynamic performance of the corrugated wing, as discussed in Ref. \cite{fujita2023dynamic} in the case of $\Rey=4000$. We quantified the Reynolds-number dependence of the mechanism. We shifted the $s^*$ axis to capture the oscillation of the LEV by setting $\xi^* = 0.15$ and considered the spatiotemporal region characterized by $0.5\leq t^*\leq1.78$ and $s^*_1\leq s^*\leq s^*_2$. The vortex behavior was characterized as

\begin{equation}
	\overline{\omega^*_z (t^*)}=\frac{1}{{s^*_2}-{s^*_1}} \int_{s^*_1}^{s^*_2} \omega^*_z (s^*, t^*) ds^*.
\label{eq:3.3}
\end{equation}

The spatial sampling region is narrower than the definition of $U_s (t^*)$ (Eq. \ref{eq:3.2}) because we focus on the single V-shaped region containing the leading edge.

Figure \ref{fig:10} (a) shows a time series of $\overline{\omega^*_z (t^*)}$ at $(\alpha, \phi, \Rey)=(1, 45^\circ, 1500)$, which succeeded in capturing the oscillation in Fig. \ref{fig:7} (b) at $\Rey=1500$, as quantified by Fourier analysis (Fig. \ref{fig:10} (b)). In this case, the highest amplitude was observed at $f=2.36$. Improvements in wing performance caused by oscillation were correlated with amplitudes observed at frequencies of $f=1.57$, $2.36$ and $3.15$. The sum of these amplitudes $A$ was used to characterize the oscillation.

The scatter plots for $\dlmean$ and $A$ are shown in Fig. \ref{fig:10} (c). The blue triangles indicate $\Rey <2000$ and the straight blue line represents the regression curve $A=10^{16.08\dlmean+0.29}$ (Fig. \ref{fig:10} (c)). The improved wing performance is strongly correlated with $A$. Therefore, for $\Rey<2000$, the oscillations of the LEV with a frequency of $1.57\leq f\leq3.15$ improve wing performance, as demonstrated in Ref. \cite{fujita2023dynamic}.

For $2000 \geq \Rey$, the scatter plot is represented by red triangles (Fig. \ref{fig:10} (c)). Independent of wing performance, $A$ maintains relatively larger values. However, the improved wing performance is correlated with $A$. The regression curve in this case exhibited a lower rate of change than that in the case of $\Rey<2000$ (Fig. \ref{fig:10} (c), red line; $A=10^{3.73\dlmean+1.35}$).

For $\Rey<2000$, $A$ has a strong correlation with $\dlmean$. Even for $\Rey \geq 2000$, there is a weak correlation with $\dlmean$; however, it is not strong. This confirms that the oscillation of the LEV at the leading edge of the corrugated wing enhanced aerodynamic performance. Vortex oscillations near the leading edge suggest that LEVs and $\lambda$-vortices alternately fit into the troughs of the corrugated wing. Consequently, stronger low-pressure regions are continuously formed within corrugations, contributing to the lift enhancement of the corrugated wing, as reported in Ref. \cite{fujita2023dynamic}. The range $\Rey \geq 2000$ corresponds to $\Reyv \geq 60$, where the vortex oscillation, vortex shedding, and activation of vortex dynamics are consistently observed.

\subsection{\label{sec:3.6} $\lambda$-vortex eruption associated with lift suppression}

\begin{figure}
	\centering{\includegraphics{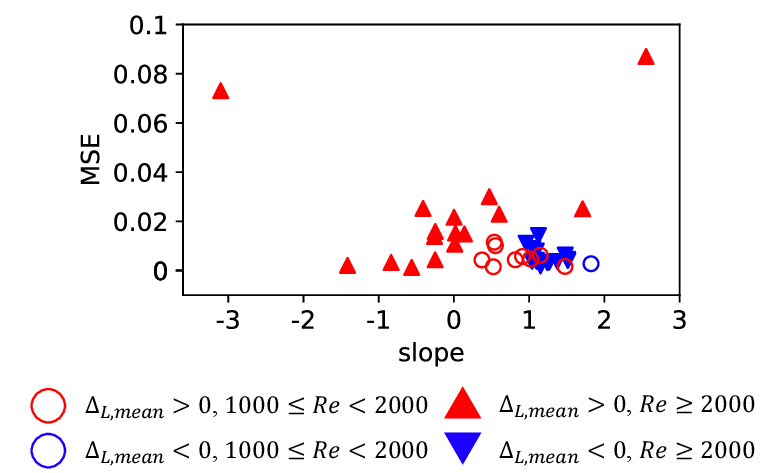}}
	\caption{\label{fig:11} Scatter plot of the slope and mean squared error for the result of pick up of the $\lambda$-vortex eruption for the corrugated wing ($\alpha=1$).}
\end{figure}

For $2000\leq \Rey \leq 4000$, where the flow around the corrugated wing is affected by the corrugated structure, the $\lambda$-vortex eruption is a factor in the decrease aerodynamic performance of the corrugated wing. To confirm this, the analysis of the $\lambda$-vortex eruption performed in our previous study \citep{fujita2023dynamic} was extended to all Reynolds numbers investigated in this study. At $\xi^* = 0.15$, points satisfying the condition that $\omega^*_z \geq 20$ is in the range $1.5\leq t^*\leq2.5$ and $0.25\leq s^*$ are selected, and a linear fit using the least-squares method is applied to that point group.

Figure \ref{fig:11} shows a scatter plot of the slope of the regression line and mean squared error (MSE). For all values of $\phi$ for $\Rey < 1000$ and $\phi=20^\circ$ and $25^\circ$ at $\Rey = 1000$, no points of $\omega^*_z$ satisfy the condition. The triangles represent corrugated-wing data in $\Rey \geq 2000$. The blue inverted triangles represent the data for cases in which $\dlmean<0$ and $2000 \leq \Rey$. Regression line clusters were characterised by a constant slope and small MSE corresponding to the $\lambda$-vortex eruption. However, when $\dlmean>0$ and $\Rey \geq 2000$ (Fig. \ref{fig:11}; red triangles), the results were scattered, which was the same trend as the results for $\Rey=4000$ \citep{fujita2023dynamic}.

The circles represent corrugated-wing data in $1000 \leq \Rey < 2000$. These points are confined to a region that overlaps with that of the inverted triangles, corresponding to a decrease in the wing performance caused by the $\lambda$-vortex eruption. However, within this Reynolds number range, many circles indicate an increase in wing performance (highlighted in red), suggesting that the $\lambda$-vortex eruption does not necessarily lead to a decrease in wing performance. Consequently, the oscillations of the LEV enhance the performance of the corrugated wing.

For $\Rey < 2000$, the $\lambda$-vortex eruption does not affect the aerodynamic performance of the corrugated wing. However, for $\Rey \geq 2000$, the $\lambda$-vortex eruption was observed when the aerodynamic performance of the corrugated wing decreased. The eruption of the $\lambda$ vortex caused by vortex interactions accelerated the release of the LEV. Consequently, the formation of low pressure on the upper surface of the wing was weakened within the analysis time, leading to a reduction in lift \citep{fujita2023dynamic}. The range $\Rey \geq 2000$ corresponded to $\Reyv \geq 60$, where the appearance of $\lambda$ vortices, vortex shedding, and activation of vortex dynamics were observed consistently.

\subsection{\label{sec:3.1*} Lift-enhancement and lift-suppression mechanisms in long-time simulations}

In this section, we examine whether the lift-enhancement and lift-suppression mechanisms identified in Ref. \cite{fujita2023dynamic}, which were originally observed during the short-time response following an impulsive start, persist in long-time simulations under fixed-wing conditions. The objective is to indicate that these mechanisms occur repeatedly for a corrugated wing placed in a uniform flow and are not restricted to transient dynamics immediately after start up.

\subsubsection{\label{sec:3.1*.1} Spatiotemporal vorticity fields}

\begin{figure}
	\centering{\includegraphics{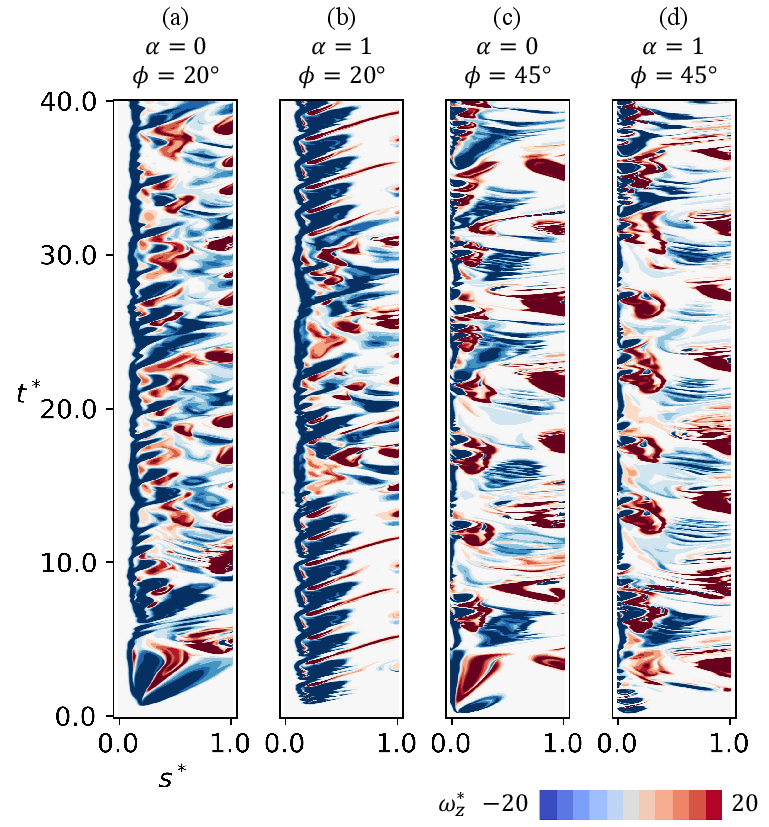}}
	\caption{\label{fig:13} Spatiotemporal vorticity distribution in a long time range ($0 \le t^* \le 40$) for the flat wing $(\alpha=0)$ and corrugated wing $(\alpha=1)$. (a) $(\alpha, \phi)=(0, 20^\circ)$. (b) $(1, 20^\circ)$. (c) $(0, 45^\circ)$. (d) $(1, 45^\circ)$. }
\end{figure}

Figure \ref{fig:13} shows spatiotemporal vorticity fields for both flat ($\alpha=0$) and corrugated wings ($\alpha=1$) at AoAs $\phi=20^\circ$ and $45^\circ$ over the time interval $0 \le t^* \le 40$.

For the flat-wing case, self-similar vortex growth was observed during the initial stage ($0 \le t^* \le 4$), irrespective of the AoA. However, this pattern was not sustained over longer periods and did not occur repeatedly. In contrast, for corrugated-wing cases, the characteristic spatiotemporal vortex patterns during the initial stage appeared frequently throughout the simulation for both AoAs considered.

For the corrugated wing at $\phi=20^\circ$, a quasiperiodic pattern was observed over the simulation time, except during the interval $t^* \in [15,30]$, when vortex structures near the trailing edge were not convected downstream. For $\phi=45^\circ$, the vortex patterns were not strictly periodic; however, vortex structures associated with lift enhancement were repeatedly observed. The characteristic duration of a single event was approximately $4$, comparable to the timescale used in the subsequent analyses.

\subsubsection{\label{sec:3.1*.2} $\lambda$-vortex eruption events and their relation to lift suppression}

\begin{figure}
	\centering{\includegraphics[width=1.\columnwidth]{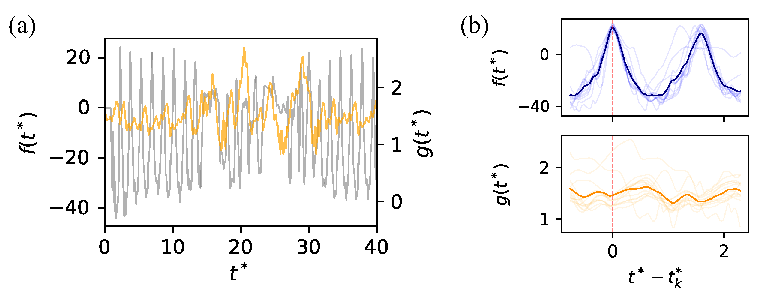}}
	\caption{\label{fig:14} Phase average of the $\lambda$-vortex-eruption event. (a) Time series of spatially averaged vorticity in the region $s^* \in [0.25, 0.375]$ on $\xi^* =0.15$, $f(t^*)$ (grey) and lift coefficient $g(t^*)$ (orange). (b) Phase-averaged waveforms of $f(t^*)$ and $g(t^*)$ obtained by aligning individual events at the peak time $t^*_k$ of $f(t^*)$. Thin and thick curves represent individual events and ensemble averages, respectively.}
\end{figure}

We focus on the corrugated wing at $\phi=20^\circ$ to demonstrate that lift suppression is associated with a $\lambda$-vortex-eruption event throughout the simulation time. We use the term vortex eruption to describe a localised and intermittent detachment of a vortex structure that forms on the wing surface; subsequently, it is expelled into the surrounding flow because of interactions with neighboring vortices. This process is distinct from the quasiperiodic vortex shedding commonly observed at sharp edges or bluff-body wakes \citep{fujita2023dynamic}.

We define $f(t^*)$ as the time series of vorticity spatially averaged over the region $s^* \in [0.25, 0.375]$ on $\xi^* =0.15$, and $g(t^*)$ as the lift coefficient, where the averaging region in $s^*$ is selected to capture the vortex eruption.

Figure \ref{fig:14} (a) shows the time series of $f(t^*)$ and $g(t^*)$. As shown in Fig. \ref{fig:13} (b), both periodic oscillations and irregular fluctuations are observed, with the latter appearing pronounced in the interval $t^* \in [15,30]$. The initial value of $f(t^*)$ is zero because vortex structures generated by the impulsive start are yet to reach the observation location at $\xi^* = 0.15$, which is away from the wing.

Individual events were extracted to statistically characterize the relationship between vortex-eruption events and lift variation. A vortex-eruption event is identified using a time interval during which
\[
f(t^*) > \omega_{\textrm{th}},
\]
where the threshold value $\omega_{\textrm{th}}(=15)$ was selected to extract significant excursions of $f(t^*)$ associated with pronounced vortex activity.

Within each event, the time when $f(t^*)$ is the maximum is denoted by $t=t^*_k$ $(k=1,2,\ldots,N_e)$. Let $T(=1.538)$ represent the characteristic time scale associated with the dominant frequency of $f(t^*)$. For each $t^*_k$, signals $f(t^*)$ and $g(t^*)$ are extracted over the interval $[t^*_k-\tfrac{1}{2}T,\,t^*_k+\tfrac{3}{2}T]$. The shifted signals $f(t^*-t^*_k)$ and $g(t^*-t^*_k)$ are subsequently superimposed with their ensemble averages, as shown in Fig. \ref{fig:14} (b). In the present dataset, a total of $N_e=13$ events are extracted.

Figure \ref{fig:14} (b) shows a phase-averaged representation of vortex-eruption events. Although event selection is based solely on the magnitude of $f(t^*)$ and includes cases in which vortex structures are highly disordered, the extracted waveforms exhibit a consistent trend. The lift coefficient does not exhibit a significant increase following the peak of $f(t^*)$. This provides clear evidence that vortex-eruption events are systematically accompanied by lift suppression under uniform flow conditions.

\subsubsection{\label{sec:3.1*.3}Vortex oscillations near the leading edge and their relation to lift enhancement}

\begin{figure}
	\centerline{\includegraphics[width=1.\columnwidth]{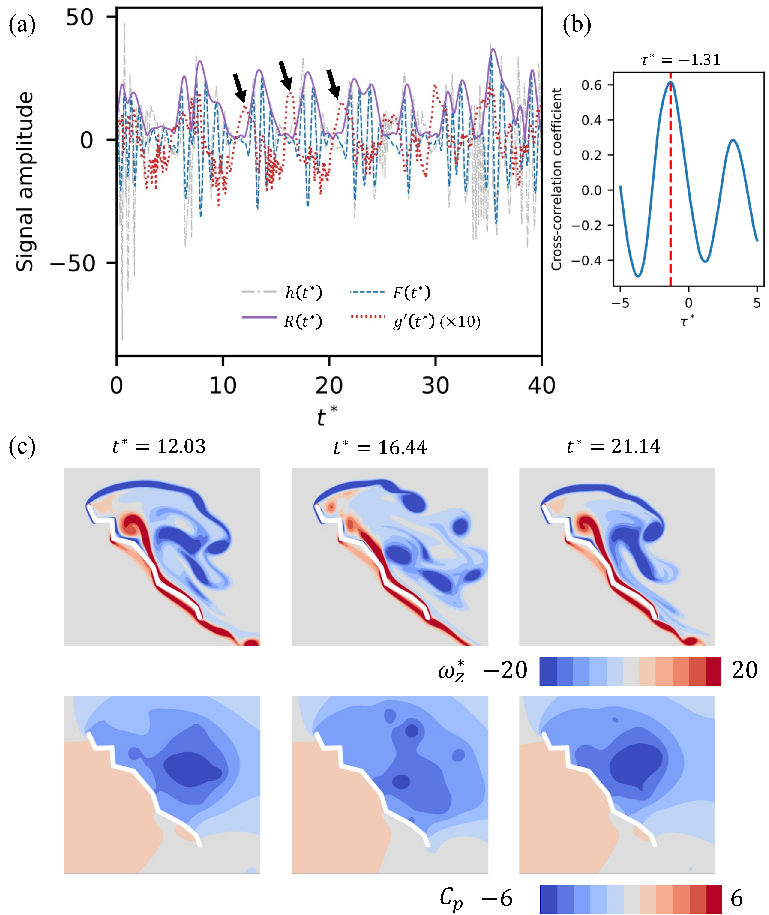}}
	\caption{\label{fig:15} (a) Time series of spatially averaged vorticity $h(t^*)$, band-pass filtered signal $F(t^*)$, and amplitude of $F(t^*)$ by Hilbert transformation $R(t^*)$. (b) Cross-correlation of $R(t^*)$ and fluctuation of the lift coefficient about its temporal mean $g'(t^*)$. (c) Snapshots of the flow field at selected times corresponding to lift peaks, as indicated by arrows in panel (a). Upper panels show the vorticity fields, and lower panels show the pressure fields.}
\end{figure}

We examine the relationship between two types of events: (i) Vortex oscillations near the leading edge at a characteristic frequency, and (ii) variations in the lift coefficient. To extract event (i), two temporal signal-processing operations are sequentially applied to the spatially averaged vorticity over the region $s^* \in [s^*_1, s^*_2]$ on $\xi^* =0.15$, $h(t^*)$. $h(t^*)$ captures the vorticity near the first concave region from the leading edge, whereas $f(t^*)$ captures the vorticity near the second concave region to detect the $\lambda$ vortex eruption.

A bandpass filter $F_{\mathrm{bp}}$ is applied over the frequency range $[\tfrac{1}{2}f_{\mathrm{p}},\, \tfrac{3}{2}f_{\mathrm{p}}]$, where $f_{\mathrm{p}}\,(=1.125)$ denotes the dominant peak frequency identified from the power spectrum. This frequency corresponds to the vortex oscillations reported in Ref. \cite{fujita2023dynamic}. The filtered signal is denoted by $F(t^*) = F_{\mathrm{bp}}[h](t^*)$. The Hilbert transform $H$ is applied to $F(t^*)$ to extract the local amplitude of the oscillatory component. The Hilbert transform of the signal $p(t)$ is defined as
\[
H(p)(t) = \frac{1}{\pi} \, \mathrm{p.v.} \int_{-\infty}^{\infty} \frac{p(\zeta)}{t - \zeta} \, \mathrm{d}\zeta,
\]
where p.v.\ represents the Cauchy principal value. We define the analytic signal $S(t^*)= F(t^*) + i F^{H}(t^*)$, where
$
F^{H}(t^*) = H(F)(t^*)
$,
and $R(t^*)=|S(t^*)|$ represents the local amplitude of the vorticity oscillation near the leading edge at the characteristic frequency.

Figure \ref{fig:15} (a) shows the signals used in the analysis, namely, $h(t^*)$, $F(t^*)$, $R(t^*)$, and the fluctuation of the lift coefficient for its temporal mean, $g'(t^*) = g(t^*)-\overline{g}$. These results confirm that $R(t^*)$ successfully extracts the local amplitude of the oscillatory component of $h(t^*)$ at the selected frequency.

Figure \ref{fig:15} (b) shows the cross-correlation between $R(t^*)$ and $g'(t^*)$ as a function of the time lag $\tau^*$. The cross-correlation reaches its maximum at $\tau^* = -1.31$, indicating that the vortex oscillation lags behind the lift peak.

This overall tendency confirms that the oscillation amplitude serves as an indicator of lift-enhancement events with systematic lag. Although Fig. \ref{fig:15} (a) shows an approximate coincidence between peaks of $R(t^*)$ and $g'(t^*)$ at $t^* \approx 1.7$ and $35$; other peaks of $R(t^*)$ occur after those of $g(t^*)$. Figure \ref{fig:15} (c) shows the instantaneous flow fields of vorticity and pressure corresponding to representative lift peaks at $t^* =12.03, 16.44$, and $21.14$. In these cases, a region of negative pressure is widely distributed over the upper surface of the wing, with strong pressure deficits observed near the trailing-edge region. This widespread low-pressure region directly contributes to the transient increase in lift.

Corresponding vortical structures consist of coherent vortices with negative vorticity, which originate from the LEV. As these vortices convect downstream, they stretch and roll up, eventually fragmenting into multiple vortical structures. Such behavior is observed during lift-enhancement events associated with vortex oscillations near the leading edge (cf.\ Figures 7 and 8 in Ref. \cite{fujita2023dynamic}).

After the lift peak is induced by the close proximity of these coherent negative-vorticity vortices, a $\lambda$ vortex develops as a consequence of the locally induced upstream flow. This upstream flow is generated through the self- and mutual induction of vortices originating from the LEV, as described by the Biot--Savart mechanism. Consequently, the development of a positive vortex emerging from the second concave region becomes clearly observable. These observations suggest a connection to statistical relationships identified by cross-correlation analysis.

\subsubsection{\label{sec:3.1*.4} Summary}
We demonstrate that the vortex-eruption and vortex-oscillation phenomena previously reported in Ref. \cite{fujita2023dynamic} are not limited to the short-time response following an impulsive start. Through long-time simulations of a corrugated wing placed in uniform flow, we showed that these vortex dynamics repeatedly occur and can therefore be considered intrinsic features of the flow-wing interaction.

Although the self-similar vortex growth in the flat-wing case is confined to the initial transient stage, the corrugated wing exhibits recurrent vortex patterns over long periods. This recurrence enables a statistical assessment of the relationship between vortex dynamics and lift generation. Vortex-eruption events are consistently associated with lift suppression, whereas vortex oscillations near the leading edge are linked to lift enhancement, albeit with a systematic time lag. These findings indicate that the lift-enhancement and lift-suppression mechanisms remain robust under uniform flow conditions.

\section{\label{sec:4}Concluding remarks}
In this study, two-dimensional direct numerical simulations were employed to evaluate the aerodynamic performance of a corrugated wing over a wide range of Reynolds numbers ($100 \leq \Rey \leq 4000$). The dependence of the lift generation on Reynolds number was investigated to clarify the operating regimes of the lift-enhancement mechanisms proposed in Ref. \cite{fujita2023dynamic}. Consequently, two distinct mechanisms, i.e., vortex oscillations associated with alternating vortex motion near the leading edge and the eruption of the $\lambda$ vortex, were confirmed to operate in different Reynolds number regimes and affect wing performance in qualitatively different ways. Long-time computations performed for representative high-Reynolds-number cases ($\Rey = 4000$) confirmed that the identified vortex dynamics and lift-generation mechanisms persisted beyond the initial transient.

For $\Rey < 1000$, the corrugated structure had a negligible effect on both the flow field and aerodynamic performance (Figs. \ref{fig:4}, \ref{fig:5}, and \ref{fig:6}) because the local Reynolds numbers inside corrugated regions were too small for secondary vortical motions to develop effectively. This regime is relevant to insects such as flies and mosquitoes flying at $\Rey \lesssim 100$ \citep{bomphrey2017smart, zhang2019role}, whose wings exhibit a weaker corrugation compared to that of dragonfly wings \citep{sudo2000wing}.

In the intermediate regime, $1000 \leq \Rey < 2000$, vortex oscillations near the leading edge emerge as a key mechanism affecting the lift. When wing performance was enhanced, oscillations with characteristic frequencies in the range $1.57 \leq f \leq 3.15$ were observed (Fig. \ref{fig:10} (c)). Although $\lambda$-vortex eruption was detected across all Reynolds numbers (Fig. \ref{fig:11}), no clear correlation between this phenomenon and lift enhancement was found in this regime. This indicates that viscous effects play a dominant role in governing the interaction between the $\lambda$ vortex and LEV. A more detailed analysis based on tracking the LEV trajectory is required and remains an important topic for future work.

When the wing performance was evaluated using the average lift coefficient over the AoA as a function of $\Rey$, $\overline{\dlmean(\Rey)}$, the maximum performance was obtained at $\Rey = 1500$, whereas higher Reynolds-number cases exhibited greater performance deviations compared with a flat wing (Fig. \ref{fig:4} (b)). This sensitivity confirms that, at $\Rey = O(10^3)$, corrugated wings can exhibit a wide range of aerodynamic responses based on the operating conditions. Such behavior highlights the potential of corrugated geometries for aerodynamic designs and applications in which performance adaptability is required.

For $\Rey \geq 2000$, the $\lambda$ vortex contributed more directly to the wing performance (Fig. \ref{fig:11}). The detailed mechanisms in this regime have been reported in Refs. \cite{fujita2023aerodynamic, fujita2023dynamic}. Furthermore, vortex oscillations were observed near the leading edge; however, oscillations within the same frequency range became more pronounced, and their correlation with lift enhancement was weaker than that observed in the intermediate Reynolds-number regime (Fig. \ref{fig:10} (c)).

The Reynolds-number-dependent flow characteristics identified in this study can be interpreted using local Reynolds number inside the V-shaped region $\Reyv$. The critical value for the onset of vortex shedding $\Reyv = O(10^2)$ is consistent with that reported in previous studies \citep{dudley2002biomechanics, miller2004vortices, ota2012lift}. Comparisons with vortex shedding from other uneven geometries, such as convex regions in two-dimensional dune models \citep{fujita2020dead}, indicate that $\Reyv$ provides a unified framework to characterize vortex dynamics induced by surface corrugation.

The results clarify the minimum Reynolds number at which corrugated wing structures become aerodynamically effective and define the Reynolds-number range over which their advantages are maintained. These findings are expected to contribute to establishing application limits and design guidelines for corrugated wings.

This study considers a single-corrugated geometry, and the shape dependency remains an important issue for future investigations. Such shape dependency has also been suggested by previous studies reporting variations in corrugation depth and leading-edge corrugation orientation in insect wings \citep{kesel2000aerodynamic, zhan2025numerical}. However, this study provides a basis for comparing different corrugated designs by introducing the local Reynolds number $\Reyv$ as a descriptor of the local flow characteristics rather than relying solely on global flow parameters or geometric features. Although the analysis was restricted to two-dimensional configurations, the results offer a foundation for future extensions to three-dimensional corrugated wings. Given the large number of design parameters involved, efficient exploration strategies such as machine-learning-based approaches are expected to play an important role in future optimization studies.

The analysis in this study focused on the relationship between vortex dynamics and lift generation in a two-dimensional corrugated wing. The objective was not to reproduce dragonfly flight or realistic flapping aerodynamics but to isolate the fundamental vortex--lift interactions induced by corrugation geometry under controlled conditions. Therefore, the simplified two-dimensional impulsive-translation setup with a rigid wing and single corrugation geometry serves not only as a limitation but also as an advantage for identifying the Reynolds-number ranges in which local corrugation-induced vortex mechanisms become active. This approach enables the examination of the aerodynamic role of wing corrugation without the additional complexity associated with three-dimensional effects, wing flexibility, and flapping kinematics.

Finally, three-dimensional effects and flapping motions should be addressed in future studies. Spanwise flow structures, wingtip vortices, and possible transitions to turbulence at higher Reynolds numbers \citep{yokoyama2013aerodynamic, gupta2023two} may modify the vortex dynamics discussed in this study. In this study, the analysis focused on flow development over a finite time interval following an impulsive start. Although long-time computations were performed for selected cases to confirm the persistence of identified vortex dynamics, it remains unclear whether three-dimensional effects would emerge over longer durations or under broader parameter variations. Accordingly, the Reynolds-number-dependent mechanisms identified should be interpreted as local physical processes associated with corrugation-induced vortex dynamics rather than as a direct model of insect flight aerodynamics. However, the two distinct vortex-driven lift enhancement mechanisms identified here, together with the systematic characterization of their effective Reynolds number ranges, provide a useful basis for future long-time and three-dimensional investigations.

Despite its two-dimensional scope, this study captures key mechanisms governing lift generation by corrugated wings within the present framework and provides quantitative insight into their Reynolds-number-dependent behavior. These results offer a solid basis for future experimental, numerical, and theoretical investigations into the aerodynamic and biological roles of corrugated wing structures, as well as their application to engineered flying systems.


\begin{acknowledgments}
This work was supported by JSPS KAKENHI (Grant Nos. 19K03671, 21H05303, 22KJ2316, 24K22861, 25K01160, and 26K17307), SECOM Science and Research Foundation, and Satake fund. The authors used generative AI tools for language editing and improving readability. All scientific content and conclusions were reviewed and verified by the authors.
\end{acknowledgments}


\bibliography{Refference.bib}

\end{document}